\documentclass[12pt]{article}

\usepackage{amsmath,amsfonts,latexsym,amssymb,amscd}
\usepackage{pslatex}
\usepackage[latin1]{inputenc}
\usepackage[T1]{fontenc}
\usepackage{verbatim,amsthm,graphics}
\usepackage{mathrsfs}

\usepackage{graphicx}

\newcommand{\mr}{{\mathbb R}}

\begin{document}


\title{ Uniqueness theorem for static wormholes  in Einstein-phantom scalar field theory }

\author{
Stoytcho Yazadjiev$^{}$\thanks{\tt yazad@phys.uni-sofia.bg}
\\ \\
{\it $ ^1$Department of Theoretical Physics, Faculty of Physics, Sofia
University} \\
{\it 5 J. Bourchier Blvd., Sofia 1164, Bulgaria} \\
{\it $ ^2$Theoretical Astrophysics, Eberhard-Karls University of T\"ubingen,}\\ {\it T\"ubingen 72076, Germany  }  }

\date{}

\maketitle

\begin{abstract}
In the present paper we prove a uniqueness theorem for the  regular static, traversable wormhole solutions to the Einstein-phantom scalar field theory
with two asymptotically flat regions (ends). We show that when  a certain condition on
the asymptotic values of the scalar field  is imposed such solutions are uniquely specified by their mass $M$ and the scalar charge $D$. 
The main arguments in the proof are based on  the positive energy theorem.
\end{abstract}


\sloppy

\section{Introduction}

Wormholes are among the most interesting objects predicted by  general relativity and the  alternative   gravitational theories \cite{MTV}. They are tunnels that connect different 
universes  or different regions of the same universe. In general the existence of wormholes requires matter with energy-momentum tensor violating the null energy condition \cite{V} -- in other words some kind of exotic matter. Today the existence of exotic matter does not sound so heretical --  what is more, the cosmological observations  suggest that the universe today is dominated by exotic matter  with negative pressure, the so-called dark energy. In light of this fact wormholes do not seem so exotic objects and could be a part of reality.  
It is also worth mentioning that in some alternative theories, like  Gauss-Bonnet-dilaton gravity  and  higher order curvature theories like f(R)
theories, wormholes have been constructed without any need of exotic matter \cite{KK}.   
      
Wormholes and black holes are rather different at first sight. However, from a mathematical point of view there are  common features between black holes and wormholes.
From a local point of view the static wormholes can be viewed as static spacetimes whose time slice contains a compact minimal surface corresponding to the wormhole throat.
The time slice of the Schwarzschild black hole, for example, is a 3-dimensional Riemannian manifold containing a compact minimal surface corresponding to the horizon and with two
asymptotoically flat ends, which is easily seen in isotropic coordinates. Therefore, from  Riemannian point of view  wormholes and black holes are very similar. 

It is well-known fact that the presence of an event horizon in certain cases, allows us to classify the asymptotically flat spacetimes only in terms of their
conserved asymptotic charges. For example, this is the case for the static vacuum Einstein
gravity and the static Einstein-Maxwell gravity. In  analogy with the black hole case it is natural to ask the following question: {\it Is it possible to classify the static wormholes 
in terms of their asymptotic charges, asymptotic values of the fields  and/or  charges and geometrical characteristics associated with their throat(s)? } 
 
In the most general case the posed problem is extremely difficult. However, in certain cases the desired classification can be achieved. The main purpose of 
this paper is to prove that  when a certain condition on the asymptotic values of the scalar field is imposed the regular\footnote{We mean wormhole solutions completely free from any singularities.} static wormholes with two asymptotically flat regions in the Einstein-phantom scalar field theory can be classified in terms of their asymptotic charges, namely the mass and the scalar charge. In different aspects the wormhole solutions  in  the Einstein-phantom scalar field theory  were intensively 
studied during the years and we refer the reader to \cite{EKK} and references therein.

\section{General definitions and equations}

The Einstein-phantom scalar field theory is given by the action

\begin{eqnarray}
{\cal S} = \frac{1}{16\pi}\int d^4x \sqrt{g^{(4)}} \left(R^{(4)} + 2g^{(4)\mu\nu}\nabla^{(4)}_{\mu}\varphi\nabla^{(4)}_{\nu}\varphi \right),
\end{eqnarray}
where $\varphi$ is the phantom scalar field, $\nabla^{(4)}_{\mu}$ and $R^{(4)}$ are the Levi-Civita connection and the Ricci scalar curvature with respect to the spacetime metric
$g^{(4)}_{\mu\nu}$. This action gives rise to the following field equations on the spacetime manifold $M^{(4)}$

\begin{eqnarray}\label{ESFE}
&&Ric^{(4)}_{\,\mu\nu}=-2\nabla^{(4)}_{\mu}\varphi\nabla^{(4)}_{\nu}\varphi, \\
&&\nabla^{(4)}_{\mu}\nabla^{(4)\mu}\varphi=0 \nonumber,
\end{eqnarray}
with $Ric^{(4)}_{\,\mu\nu}$ being the Ricci tensor. 

We shall  focus on {\it  static spacetimes in the strict sense} which means spacetimes with a
Killing vector field  $\xi$ which is everywhere timelike. For such spacetimes  there exists a smooth 3-dimensional Riemannian manifold $(M^{(3)}, g^{(3)}_{ij})$ and
a smooth lapse function $N: M^{(3)} \to {\mr}^{+}$ such that

\begin{eqnarray}
M^{(4)}= {\mr} \times M^{(3)}, \,\,\, g^{(4)}_{\mu\nu} dx^\mu dx^\nu= - N^2 dt^2 + g^{(3)}_{ij}dx^i dx^j.
\end{eqnarray}
In addition to the metric staticity we  have to define the scalar field staticity. The scalar field is called static if ${\cal L}_{\xi}\varphi= 0$ where ${\cal L}_{\xi}$ is
the Lie derivative along the Killing field $\xi=\frac{\partial}{\partial t}$. We note that both notions of staticity are consistent since the Ricci
1-form $Ric^{(4)}[\xi]=\xi^{\mu}Ric^{(4)}_{\,\mu\nu}\,dx^{\nu}$ is zero due to the field equations and the fact that $\xi^{\mu}\nabla^{(4)}_{\mu}\varphi={\cal L}_{\xi}\varphi= 0$.

In the present paper we are interested in the wormhole solutions to the field equations (\ref{ESFE}) with two asymptotically flat ends. We shall adopt the following formal definition of a static, asymptotically flat wormhole  solution with two ends.

\medskip
\noindent

{\it A solution to the field equations (\ref{ESFE}) is said to be  a static, asymptotically flat traversable  wormhole solution if the following conditions are satisfied:

\medskip
\noindent	
	
1. The spacetime is strictly static.

\medskip
\noindent

2. The Riemannian manifold $(M^{(3)},g^{(3)})$ is complete.  

\medskip
\noindent

3.  For some compact set $K$, $M^{(3)}-K$ consists of two ends $End_{+}$ and $End_{-}$  such that each end is diffeomorphic to  ${\mr}^{3} \backslash \bar{B}$ where $\bar{B}$ is the closed unit ball centered at the origin in ${\mr}^3$, and with the following asymptotic behavior of the 3-metric, the lapse function  and the scalar field  }

\begin{eqnarray}
g^{(3)}_{ij}= N^{-2}_{\pm}\left(1 + \frac{2M_{\pm}}{r}\right)\delta_{ij} + {\cal O}(r^{-2}), \;\;\; N= N_{\pm}\left(1 - \frac{M_{\pm}}{r}\right) + {\cal O}(r^{-2}), \;\; \varphi =\varphi_{\pm} - \frac{D_{\pm}}{r} + {\cal O}(r^{-2}),
\end{eqnarray}
{\it with respect to the standard radial coordinate $r$ of $\mr^3$, where $\delta_{ij}$ is the standard flat metric on $\mr^3$}. 

\medskip
\noindent

Here $N_{\pm}>0$, $M_{\pm}$, $\varphi_{\pm}\ne 0$ and $D_{\pm}\ne 0$ are constants.  $M_{\pm}$ and $D_{\pm}$ are the total (ADM) mass and the scalar charge of the corresponding end $End_{\pm}$. We  chose $N_{-}\le N_{+}$. 

The conditions of the wormhole definition are quite natural from a physical point of view. The first condition means that there are no horizons present. The second condition
means that  the geometry of the time slices is free from singularities. The third condition is just our standard notion of asymptotically flat regions. 

The dimensionally reduced static Einstein-scalar field equations are the following

\begin{eqnarray}\label{DRFE}
&&Ric^{(3)}_{ij} = N^{-1} \nabla^{(3)}_i \nabla^{(3)}_j N - 2 \nabla^{(3)}_i\varphi \nabla^{(3)}_j\varphi ,\nonumber \\
&&\nabla^{(3)}_i \nabla^{(3)i}N=\Delta^{(3)}N=0, \\
&&\nabla^{(3)}_i\left(N \nabla^{(3)i}\varphi\right)=0 , \nonumber
\end{eqnarray}
where $\nabla^{(3)}_i$ and $Ric^{(3)}_{ij}$ are the Levi-Civita connection and the Ricci tensor with respect to the metric $g^{(3)}_{ij}$.

By the maximum principle for harmonic functions and by the asymptotic behavior of $N$  we obtain that the values of $N$ on $M^{(3)}$
obey

\begin{eqnarray}\label{INEN}
N_{-}\le N\le N_{+},
\end{eqnarray}
as  equality is satisfied only when $N$ is constant on $M^{(3)}$ or equivalently when $M_{+}=M_{-}=0$.  Without loss of generality we can impose

\begin{eqnarray}
 N_{+}N_{-}=1 
\end{eqnarray}

and 

\begin{eqnarray}
\varphi_{+}=-\varphi_{-}.  
\end{eqnarray}
These normalizations are convenient and allow us to get rid of the arbitrary constant that can be added to the scalar field, $\varphi \to \varphi + const$.

Using again the maximum principle for elliptic partial  differential 
equations and the asymptotic behavior of $\varphi$ we conclude that on   $M^{(3)}$ we have 

\begin{eqnarray}\label{INQPHI}
\varphi_{-}< \varphi <\varphi_{+} .
\end{eqnarray}

Let us note that we must have $\varphi_{+}=-\varphi_{-}\ne 0$ otherwise the scalar field woud be trivial.

Integrating $\Delta^{(3)}N=0$ on   $M^{(3)}$ by applying the Gauss theorem and taking into account the asymptotic behavior of $N$ we obtain 

\begin{eqnarray}\label{MPM}
M_{-} + M_{+}=0. 
\end{eqnarray}

In the same way the integration of $\nabla^{(3)}_i\left(N \nabla^{(3)i}\varphi\right)=0$ on $M^{(3)}$ gives

\begin{eqnarray}\label{QPM} 
D_{-} + D_{+}=0. 
\end{eqnarray}

It is convenient to introduce $M=|M_{+}|=|M_{-}|$ and $D=|D_{-}|=|D_{+}|$ and then using (\ref{MPM}) and (\ref{QPM}) we have $M_{+}/D_{+}=M_{-}/D_{-}=M/D$. {\it We will refer to 
$M$ and $D$ as the mass and the scalar charge of the wormhole.}

As a consequence of the dimensionally reduced field equations for $N$ and $\varphi$, it is not difficult  to show that the current $J_{i}=\varphi\nabla^{(3)}_i N - N\ln(N)\nabla^{(3)}_i\varphi$ is conserved, i.e. $\nabla^{(3)}_i J^i=0$. Integrating $\nabla^{(3)}_i J^i=0$ on $M^{(3)}$ and using the asymptotic behavior of $N$ and $\varphi$
one finds 

\begin{eqnarray}\label{FDINF}
\ln(N_{\pm})= \frac{M}{D}\varphi_{\pm} .
\end{eqnarray}

By using the equation for $\varphi$ in (\ref{DRFE}) it is not difficult to show that  

\begin{eqnarray}
&&\frac{1}{4\pi}\int_{M^{(3)}} N \nabla^{(3)}_i\varphi \nabla^{(3) i}\varphi \sqrt{g^{(3)}} d^3x= \frac{1}{4\pi} \int_{M^{(3)}}   \nabla^{(3)}_i\left(\varphi N \nabla^{(3) i}\varphi\right) \sqrt{g^{(3)}} d^3x 
\\
&&= \frac{1}{4\pi}\int_{S^{2+}_\infty} \varphi N   \nabla^{(3)}_i\varphi d^2\Sigma^{i} +  \frac{1}{4\pi} \int_{S^{2-}_\infty} \varphi N   \nabla^{(3)}_i\varphi d^2\Sigma^{i}=
\varphi_{+}D_{+} + \varphi_{-}D_{-}= 2 \varphi_{+}D_{+}. \nonumber 
\end{eqnarray}
Hence we obtain an important inequality which will be used in the proof of the main theorem, namely

\begin{eqnarray}\label{PP}
\varphi_{+}D_{+}=\varphi_{-}D_{-}>0.
\end{eqnarray}

The most famous static wormhole solution to the field equations (\ref{ESFE}) is the Ellis-Bronnikov solution \cite{EB}. This solution is spherically symmetric and 
its explicit form in our notations  is the following

\begin{eqnarray}
&&ds^2 = - \exp{\left[\frac{2M}{\sqrt{D^2-M^2}}\arctan(\frac{l}{\sqrt{D^2-M^2}})\right]} dt^2  \nonumber \\ &&+ \exp{\left[-\frac{2M}{\sqrt{D^2-M^2}}\arctan(\frac{l}{\sqrt{D^2-M^2}})\right]}\left[dl^2 + (l^2 + D^2 -M^2)(d\theta^2 + \sin^2\theta d\phi^2) \right], \\
&&\varphi= \frac{D}{\sqrt{D^2-M^2}} \arctan(\frac{l}{\sqrt{D^2-M^2}}),  
\end{eqnarray}
where  $l\in (-\infty,+\infty)$ and $D^2>M^2$. The asymptotic values of the scalar field and the lapse function are correspondingly   

\begin{eqnarray}
\varphi_{\pm}= \pm \frac{D}{\sqrt{D^2-M^2}} \frac{\pi}{2}, \,\,\, \; \; N_{\pm}= \exp{\left[\pm \frac{M}{\sqrt{D^2-M^2}} \frac{\pi}{2}\right]}.
\end{eqnarray}
and obviously satisfy (\ref{FDINF}) and (\ref{PP}).

\section{ Uniqueness theorem }

The natural conjecture is that the Ellis-Bronnikov solution is the unique static asympototically flat wormhole solution with given mass $M$, scalar charge $D$ and asymptotic value(s) of the scalar field $\varphi_{+}= - \varphi_{-}=\frac{\pi}{2}\frac{D}{\sqrt{D^2-M^2}}$.  We will prove a more general result by weakening the last condition. Instead of imposing
the rigid condition  $\varphi_{+}= - \varphi_{-}=\frac{\pi}{2}\frac{D}{\sqrt{D^2-M^2}}$ we shall restrict the  asymptotic value of the scalar field to the interval 
$0<\sqrt{1-\frac{M^2}{D^2}} \varphi_{+} \le \frac{\pi}{2}$.

The following theorem is the main result of this paper:

\medskip
\noindent

{\bf Theorem} {\it The mass $M$  and the scalar charge $D$  of the static, asymptotically flat traversable wormhole solutions to  the  Einstein-phantom scalar field equations
satisfy  the inequality $M^2<D^2$. Moreover, there can be only one static, asymptotically flat traversable   wormhole spacetime $(M^{(4)}, g^{(4)},\varphi)$,  
with given mass $M$ and scalar charge $D$, and asymptotic value $\varphi_{+}$ of the scalar field in the interval $0<\sqrt{1-\frac{M^2}{D^2}} \varphi_{+} \le \frac{\pi}{2}$ and 
it  is isometric to the Ellis-Bronnikov wormhole spacetime. }

\medskip
\noindent

{\bf Proof:} The strategy of the proof is the following. First we shall prove that there is a functional dependence between the lapse function $N$ and the scalar field $\varphi$.
Then using appropriate conformal transformations we shall reduce the problem to problems where we can use  the positive energy theorem and more precisely its Riemannian version
\cite{SYB}.  

The first step is to show that 

\begin{eqnarray}\label{FD}
\ln(N)=\frac{M}{D}\varphi.
\end{eqnarray}

For this purpose let us  consider the divergence identity

\begin{eqnarray}
N^{-1}\left(\omega_{i} \,\omega^{i}\right)= \nabla^{(3)}_i \left[\left(\frac{M}{D}\varphi - \ln(N)\right)\omega^{i}\right],
\end{eqnarray}
where

\begin{eqnarray}
\omega_{i}=\frac{M}{D} N\nabla^{(3)}_i\varphi -  \nabla^{(3)}_i N.
\end{eqnarray}
As one can verify this identity is a consequence of the field equations for the lapse function and the scalar field. By applying the Gauss theorem to the above identity we obtain

\begin{eqnarray}
\int_{M^{(3)}} N^{-1}\left(\omega_{i} \,\omega^{i}\right) \sqrt{g^{(3)}}d^3x =0 ,
\end{eqnarray}
where we have taken into account the asymptotic behavior of $N$ and $\varphi$ as well as (\ref{FDINF}) in the evaluation of the surface integrals. Since $N>0$ on $M^{(3)}$
we conclude that $\omega_{i}= \frac{M}{D} N\nabla^{(3)}_i\varphi -  \nabla^{(3)}_i N=0$ on $M^{(3)}$. Therefore we obtain $\ln(N)= \frac{M}{D}\varphi + C$ with $C$  being a constant.
From  Eq.(\ref{FDINF}) we find that $C=0$ which proves (\ref{FD}).

Further we  consider the 3-metric $h_{ij}$ on $M^{(3)}$ defined by the conformal transformation 

\begin{eqnarray}\label{hmetric}
h_{ij}=N^{2}g^{(3)}_{ij}.
\end{eqnarray}

In terms of the new metric the dimensionally reduced  equations become

\begin{eqnarray}\label{HDRFE}
&&R(h)_{ij}= 2D_{i}\ln(N)D_{j}\ln(N) - 2 D_{i}\varphi D_{j}\varphi ,\nonumber\\
&&D_{i}D^{i}\ln(N) = 0, \\
&&D_{i}D^{i}\varphi = 0, \nonumber
\end{eqnarray}
where $D_{i}$ and $R(h)_{ij}$ are the Levi-Civita connection and the Ricci tensor with respect to $h_{ij}$, respectively.  Taking into account the functional dependence
$\ln(N)=\frac{M}{D}\varphi$ the  equations (\ref{HDRFE}) can be cast in the form

\begin{eqnarray}\label{EFFE}
&&R(h)_{ij}= -2\left(1- \frac{M^2}{D^2}\right)D_{i}\varphi D_{j}\varphi. \\
&&D_{i}D^{i}\varphi = 0. \nonumber
\end{eqnarray}

The three-dimensional Riemannian manifold $(M^{(3)},h_{ij})$ is a complete asymptotically flat manifold with two ends of vanishing mass, $M^{h}_{\pm}=0$. 
The fact that the metric $h_{ij}$ is complete follows directly from the definition of $h_{ij}$, the fact that $g^{(3)}_{ij}$ is complete and the inequality (\ref{INEN}).
The fact that total mass of each end is zero follows from the asymptotic behavior of $h_{ij}$, namely 

\begin{eqnarray}
h_{ij}= \delta_{ij} + O(r^{-2}),
\end{eqnarray}     
which can be easily obtained from the asymptotic behavior of  $g^{(3)}_{ij}$  and $N$. 

Now let us assume for a moment that $M^2\ge D^2$. Then we have an asymptotically flat Riemannian manifold  $(M^{(3)},h_{ij})$  which is complete, with a non-negative scalar curvature (as can be seen from   (\ref{EFFE})) and with zero total mass for each of its ends. From the rigidity of the positive energy theorem \cite{SYB}  it follows that   $(M^{(3)},h_{ij})$ is isometric to $(\mr^3, \delta_{ij})$. So we obtain a contradiction due to our assumption that  $M^2\ge D^2$. Therefore we conclude that for wormhole solutions  the mass and the scalar charge have to satisfy
the inequality 

\begin{eqnarray}\label{MCHINQ}
M^2< D^2. 
\end{eqnarray}

Having proven that $M^2< D^2$ we can introduce a new scalar field $\lambda$, defined by 

\begin{eqnarray}\label{LambdaDEF}
 \lambda=\sqrt{1-\frac{M^2}{D^2}}\varphi. 
\end{eqnarray}

In terms of $\lambda$ equations (\ref{EFFE}) take the form

\begin{eqnarray}\label{EFFEL}
&&R(h)_{ij}= -2D_{i}\lambda D_{j}\lambda, \\
&&D_{i}D^{i}\lambda = 0, \nonumber
\end{eqnarray}
with  $0<\lambda_{+}=-\lambda_{-}\le\frac{\pi}{2}$.

The next step is to consider the following  metric 

\begin{eqnarray}
\gamma_{ij}= \Omega^2 h_{ij}  
\end{eqnarray} 
where  the conformal factor $\Omega^2$ is given by 

\begin{eqnarray}
	\Omega^2 = \frac{\sin^4(\frac{\lambda + \lambda_{+} }{2})}{\sin^4(\lambda_{+})} .
\end{eqnarray}

Using that  the scalar curvature $R(\gamma)$ of $\gamma_{ij}$ is given by  

\begin{eqnarray} 
R(\gamma)= \Omega^{-2} R(h) - 4\Omega^{-3}D_{i}D^i\Omega + 2\Omega^{-4}D_i\Omega D^i\Omega   
\end{eqnarray}  
and the equations (\ref{EFFEL}) we obtain that $\gamma_{ij}$ has zero scalar curvature, $R(\gamma)=0$. The end $End_{-}$ can be compactified \cite{BMUA} by adding in a point $\infty$ at infinity 
since 

\begin{eqnarray}
{\gamma}_{ij}=\Omega^2 h_{ij}= \frac{(D^2-M^2)^2}{16 \sin^4(\lambda_{+}) r^4} \delta_{ij} + O(1/r^6)
\end{eqnarray}
as $r\to \infty$ in the standard  asymptotic coordinates for $End_{-}$. Performing the coordinate transformation ${\tilde x}^i= x^i/r^2$ we find that in these new coordinates we have

\begin{eqnarray}
{\gamma}(\frac{\partial}{\partial{\tilde x}^i},\frac{\partial}{\partial{\tilde x}^j})= \frac{(D^2-M^2)^2}{16 \sin^4(\lambda_{+})} \delta_{ij} + O({\tilde r }^2)
\end{eqnarray} 
as ${\tilde r} \to 0$ where ${\tilde r}^2=\delta_{ij}{\tilde x}^i {\tilde x}^j$. This indeed shows that we can add in a point $\infty$ at ${\tilde r}=0$  so that the constructed in this way manifold 
${\tilde M^{(3)}}=M^{(3)} \cup \infty$ is (sufficiently) regular. By construction ${\tilde M^{(3)}}=M^{(3)} \cup \infty$ is geodesically complete and with only one end, namely $End_{+}$.

The Riemannian manifold $({\tilde M^{(3)}},\gamma_{ij})$ is geodesically complete, scalar flat manifold  with one asymptotically flat end and according to the positive energy theorem
\cite{SYB} its total mass ${\tilde M}^{\gamma}$ with respect to the metric $\gamma_{ij}$ must be non-negative,  ${\tilde M}^{\gamma}\ge 0$. The mass ${\tilde M}^{\gamma}$ can be found from the asymptotic behaviour of $\gamma_{ij}$, namely 

\begin{eqnarray}
\gamma_{ij}= \left(1- \frac{2D_{+}\cot(\lambda_{+})\sqrt{1-\frac{M^2}{D^2}} }{r}\right)\delta_{ij} + O(r^{-2})
\end{eqnarray}    
as $r\to \infty$. Hence we have ${\tilde M}^{\gamma}= -2D_{+}\cot(\lambda_{+})\sqrt{1-\frac{M^2}{D^2}}$. From (\ref{PP}) it follows that  ${\tilde M}^{\gamma}\le 0$
as the equality is saturated only for $\lambda_{+}=\frac{\pi}{2}$. Therefore we conclude that  ${\tilde M}^{\gamma}=0$ and $\lambda_{+}=\frac{\pi}{2}$. 

Summarizing, $({\tilde M^{(3)}},\gamma_{ij})$ is geodesically complete, scalar flat Riemannian
manifold with one asymptotically flat end of vanishing total mass. Then the rigidity of the positive energy  theorem \cite{SYB}  guarantees that $({\tilde M^{(3)}},\gamma_{ij})$ is isometric to 
$(\mr^3, \delta_{ij})$. This in turn means that the metrics $h_{ij}$ and $g^{(3)}_{ij}$  are conformally flat and $M^{(3)}$ is diffeomorphic to $\mr^3/\{0\} $.

The theorem now follows  by straightforward integration of the 
field equations (\ref{EFFEL}) in spherical coordinates. More precisely we have to integrate   (\ref{EFFEL})  for the metric

\begin{eqnarray}
h_{ij}dx^idx^j= \sin^{-4}(\frac{\lambda }{2} + \frac{\pi }{4} )(dR^2 + R^2 d\theta^2 + R^2\sin^2\theta d\phi^2)
\end{eqnarray} 
with the  asymptotic conditions for $\lambda$ described in the second section. It turns out more convenient to write down the field equations  (\ref{EFFEL}) in
terms of the new scalar field $\sigma=\cot(\frac{\lambda}{2} + \frac{\pi}{4} )$ and the flat metric $\gamma_{ij}$. Taking into account the conformal properties of the Ricci tensor
one can show that  equations  (\ref{EFFEL}) are equivalent to the following system    

\begin{eqnarray}\label{SigmaEQ}
&&-\sigma D^{\gamma}_{i}D^{\gamma}_{j}\sigma + 3 D^{\gamma}_{i}\sigma D^{\gamma}_{j}\sigma - \gamma_{ij} \gamma^{mn}D^{\gamma}_{m}\sigma D^{\gamma}_{n}\sigma=0,\\
&&\gamma^{ij}D^{\gamma}_{i}D^{\gamma}_{j}\sigma=0,\nonumber
\end{eqnarray}
where $D^{\gamma}_{i}$ is the covariant derivative with respect to the flat metric $\gamma_{ij}$. The asymptotic behavior of $\lambda$ translated for $\sigma$  
is 

\begin{eqnarray}
&&\sigma= \frac{\sqrt{D^2-M^2}}{2R}  + O(R^{-2}), \; R\to \infty, \\
&&\sigma=  \frac{\sqrt{D^2-M^2}}{2R} + O(1), \; R \to 0. 
\end{eqnarray} 

The equation for $\sigma$ is just the standard Laplace equation on  $(\mr^3/\{0\}, \, \delta_{ij})$ and  its unique solution satisfying the above asymptotics is 

\begin{eqnarray}
\sigma= \frac{\sqrt{D^2-M^2}}{2R}. 
\end{eqnarray} 
It is not difficult  to check that this solution satisfies also the remaining equations of (\ref{SigmaEQ}).  Hence we find\footnote{We use also that $\arctan(x)=\frac{\pi}{2}- \arctan(\frac{1}{x})$ for $x>0$.} 

\begin{eqnarray}
&&h_{ij}dx^idx^j= \left(1+ \frac{D^2-M^2}{4R^2}\right)^2 \left(dR^2 + R^2 d\theta^2 + R^2\sin^2\theta d\phi^2\right),\\\
&&\lambda= 2\arctan(\frac{2R}{\sqrt{D^2-M^2}}) -\frac{\pi}{2}, 
\end{eqnarray}
where $R\in(0,+\infty)$. The end $End_{+}$ corresponds to $R\to \infty$ while the end $End_{-}$ corresponds to $R\to 0$. 
The solution can be presented in the more familiar coordinate $l\in(-\infty,\infty)$ by  the coordinate transformation\footnote{One should also use that $2\arctan(\frac{l+ \sqrt{l^2+a^2}}{a})=\frac{\pi}{2} + \arctan(\frac{l}{a})$.} $l= R- \frac{D^2-M^2}{4R}$, namely

\begin{eqnarray}
&&h_{ij}dx^idx^j= dl^2 + (l^2 + D^2- M^2)(d\theta^2 + \sin^2\theta d\phi^2),\\
&&\lambda=\arctan(\frac{l}{\sqrt{D^2-M^2}}). 
\end{eqnarray}	

Taking into account (\ref{LambdaDEF}), (\ref{FDINF}) and (\ref{hmetric}) we can find $\varphi$, $N$ and $g^{(3)}_{ij}$ and they are exactly the scalar field, the lapse function 
and the 3-metric of the Ellis-Bronnikov wormhole solution. This completes the proof of the theorem.

The following comments are in order. Strictly speaking the uniqueness is proven up to the trivial non-uniqueness related to the sign of the scalar field -- both $+\varphi$ and 
$-\varphi$  with the  three-metric $h_{ij}$ unchanged are solutions to the field equations. This trivial non-uniqueness corresponds to the interchange of the asymptotic regions of 
the wormhole. If we abandon our normalization $\varphi_{+}=-\varphi_{-}$ the condition  $0<\sqrt{1-\frac{M^2}{D^2}} \varphi_{+} \le \frac{\pi}{2}$ changes to  $0<\sqrt{1-\frac{M^2}{D^2}} (\varphi_{+} -\varphi_{-})\le \pi$ and the very proof gives $\sqrt{1-\frac{M^2}{D^2}} (\varphi_{+} -\varphi_{-})= \pi$.

\section{Conclusion}

In the present paper we proved a uniqueness  theorem for  the  static, traversable wormhole solutions to the Einstein-phantom scalar field theory
with two asymptotically flat ends and completely free from singularities. To the best of our knowledge  this is the first classification theorem for traversable  wormhole solutions.  Our theorem can be extended to more general equations, for example to the case of the Einstein-Maxwell-(phantom) scalar equations. Another direction of investigation is 
to consider wormhole solutions with arbitrary (but finite) number of asymptotically flat ends. All these problems will be discussed in future publications \cite{LNY}.   

\medskip
\noindent

{\bf Acknowledgements:}  The partial support by Bulgarian NSF grant DFNI T02/6,  Sofia University Research Grant  80.10-30/2017 
and  the COST Action CA16104 is  gratefully acknowledged.

\end{document}